# Identify Main-sequence Binaries from the Chinese Space Station Telescope Survey with Machine Learning. II. Based on Gaia and GALEX

Jia-jia Li[1,2], Jian-ping Xiong[1], Zhi-jia Tian[3], Chao Liu[2,4], Zhan-wen Han[1,2,5,6], and Xue-fei Chen[1,2,5,6]
[1] Yunnan Observatories, Chinese Academy of Sciences, 396 Yangfangwang, Guandu District, Kunming, 650216, People's Republic of China; lijiajia@ynao.ac.cn, cxf@ynao.ac.cn
[2] School of Astronomy and Space Science, University of Chinese Academy of Sciences, Beijing, 100049, People's Republic of China
[3] Yunnan University, Kunming 650091, People's Republic of China
[4] Key Laboratory of Space Astronomy and Technology, National Astronomical Observatories, Chinese Academy of Sciences, Beijing, 100101, People's Republic of China
[5] Center for Astronomical Mega-Science, Chinese Academy of Sciences, Beijing 100012, People's Republic of China
[6] International Centre of Supernovae, Yunnan Key Laboratory, Kunming 650216, People's Republic of China
Received 2024 December 19; revised 2025 February 19; accepted 2025 February 20; published 2025 March 21

## Abstract

The statistical characteristics of double main-sequence (MS) binaries are essential for investigating star formation, binary evolution, and population synthesis. Our previous study proposed a machine learning-based method to identify MS binaries from MS single stars using mock data from the Chinese Space Station Telescope (CSST). We further utilized detection efficiencies and an empirical mass ratio distribution to estimate the binary fraction within the sample. To further validate the effectiveness of this method, we conducted a more realistic sample simulation, incorporating additional factors such as metallicity, extinction, and photometric errors from CSST simulations. The detection efficiency for binaries with mass ratios between 0.2 and 0.7 reached over 80%. We performed a detailed observational validation using the data selected from the Gaia Sky Survey and Galaxy Evolution Explorer. The detection efficiency for MS binaries in the observed sample was 65%. The binary fraction can be inferred with high precision for a set of observed samples, based on accurate empirical mass ratio distribution.

*Unified Astronomy Thesaurus concepts:* Binary stars (154)

## 1. Introduction

Research shows that more than 50% of stars in the Milky Way are part of binary or multiple star systems (W. D. Heintz 1969; H. A. Abt & S. G. Levy 1976; A. Duquennoy & M. Mayor 1991; H. Sana et al. 2012; M. Moe & R. Di Stefano 2017). Understanding binary systems is essential for studying stellar evolution and cosmic phenomena (Z.-W. Han et al. 2020). Binaries are key to measuring the dynamic masses and radii of stars (S. A. Tjemkes et al. 1986; H. Hensberge et al. 2000; A. L. Kraus et al. 2011; J. Xiong et al. 2023), which is crucial for refining stellar evolution models (G. Burbidge et al. 1980; C. Conroy et al. 2009; X. Chen et al. 2024). Binary systems serve as a key to understanding the formation mechanisms of other intriguing objects, such as compact binaries, supernovae, gamma-ray bursts, X-ray binaries, pulsars, cataclysmic variables, etc. (H. Sana et al. 2012; Z.-W. Han et al. 2020). Among these objects, Type Ia supernovae serve as standard candles for measuring cosmic distances, and their classical progenitor models are also binary systems (B. Wang & Z. Han 2012). And the double neutron stars, double black holes, black hole-neutron binaries and double dwarfs mergers, sources of gravitational waves, validate general relativity and reveal insights into extreme gravitational fields (LIGO Scientific Collaboration et al. 2015; P. Amaro-Seoane et al. 2023). Additionally, the advent of large sample surveys, such as those from Gaia, Pan-STARRS, and other large-scale sky surveys, has opened up exciting opportunities for precise distance measurements of binary stars, particularly through the study of eclipsing binaries (D. Graczyk et al. 2017; G. Pietrzyński et al. 2019).

Binary population synthesis (BPS) is a powerful computational method used to simulate the evolution of binary populations, enabling the study of double black holes (V. M. Lipunov et al. 1997; E. De Donder & D. Vanbeveren 1998), double neutron stars (S. F. Portegies Zwart & L. R. Yungelson 1998; K. Belczynski et al. 2002), double white dwarf stars (Z. Han 1998), the progenitors of Ia type stars (B. Wang & Z. Han 2012), hot subdwarfs (Z. Han et al. 2003), symbiotic stars (G. Lü et al. 2006), and blue stragglers (X. Chen & Z. Han 2008). The key inputs to the BPS method include the initial masses, binary fraction, orbital parameters, metallicity, and mass transfer efficiency. Among these, the binary fraction plays a critical role, as it significantly influences the evolutionary pathways, observational signatures, and the occurrence rates of astrophysical events such as supernovae and gravitational wave sources. Ensuring an accurate representation of the binary fraction is essential for improving the alignment of model predictions with observational data (M. U. Kruckow 2018; E. R. Stanway & J. J. Eldridge 2023).

Large surveys, such as the Sloan Digital Sky Survey (SDSS), Gaia, APOGEE, LAMOST, etc., have played an important role in the study of the statistical properties of binary populations. These surveys have contributed a wealth of observational data, thereby driving numerous related research initiatives and advancing our understanding of binary fraction. (S. Gao et al. 2014; H. Yuan et al. 2015; J. J. Andrews et al. 2017; C. Badenes et al. 2018; K. El-Badry & H.-W. Rix 2019; C. N. Mazzola et al. 2020; H.-C. Hwang et al. 2022; Y. Guo et al. 2022b, 2022a). The binary fraction is linked to several stellar properties (M. Moe & R. Di Stefano 2017). Higher mass stars are observed to be more likely to be in binary systems, low-mass stars exhibit a binary fraction of approximately 20%–30% (X. Delfosse et al. 2004; P. R. Allen 2007), solar-mass stars around 40%–50% (D. Raghavan et al. 2010; J. Southworth 2021), and O- and B-type stars up to 70%–90%, often in close binaries (R. G. Izzard et al. 2018; X. Chen et al. 2024). Moreover, the binary fraction is correlated with metallicity, with higher fractions in metal-rich stars, and with age, as younger stars tend to have a higher binary fraction, which decreases over time (S. Gao et al. 2014; K. El-Badry et al. 2018).

Accurate identification of binary stars is crucial for calculating the binary fraction. The methods for identifying binary star systems depend on the type of binary, their separation, and the observational techniques available. Common identification methods include spectral analysis (e.g., B. Zhang et al. 2022; J. Liu et al. 2024; M. Kovalev et al. 2024), photometric variations (e.g., B. Kirk et al. 2016; X. Chen et al. 2020; A. Prša et al. 2022), and astrometry (e.g., N. Mowlavi et al. 2023). In addition, researchers have utilized the characteristic of binary sequences appearing brighter and redder than the main sequence (MS) on the color–magnitude diagram (CMD) and color–color diagram (CCD) to discover binaries within star clusters (K. El-Badry et al. 2018; L. Li et al. 2020; A. M. Price-Whelan et al. 2020). When the mass ratio is 1, the binary is 0.75 mag brighter than its primary star (J. Hurley & C. A. Tout 1998). However, this method of identifying binary stars is highly limited to applications within cluster environments and cannot be effectively applied to regions where single and binary stars overlap. In the current era of astronomical big data, with an explosion of all-sky observational data from various survey projects, it is essential to find faster and more efficient methods for identifying binary stars. Researchers have already employed machine learning techniques to identify quasars (B. Liu & R. Bordoloi 2021), neutron star binaries (V. S. Pérez-Dìaz et al. 2024), clusters (G. Pérez et al. 2021), and binaries (S.-Y. Lan et al. 2022; J.-j. Li et al. 2024, hereinafter referred to as Paper I) by using the multiband photometric data.

The Chinese Space Station Telescope (CSST) is a major astronomical project within China's space program. CSST, a 2 m optical/ultraviolet space telescope, is scheduled for launch around 2026. It will focus on high-precision photometric and slitless spectroscopic observations, conducting large-scale sky surveys. During its entire survey period, the CSST will cover approximately 1.7 deg$^2$ of the sky, which accounts for about 40% of the total sky area. CSST offers broader observational capabilities compared to the Hubble Space Telescope (H. Zhan 2011), significantly enhancing both efficiency and coverage. Its primary scientific goal is to collect vast amounts of astronomical data through multiband photometry. It is equipped with a range of filters covering seven bands, from ultraviolet to visible light (near-UV (NUV): 2520–3210 Å, $u$: 3210–4010 Å, $g$: 4010–5470 Å, $r$: 5470–6920 Å, $i$: 6920–8420 Å, $z$: 8420–10800 Å, and $y$: 9270–10800 Å), enabling simultaneous measurements of celestial objects across multiple wavelengths (Y. Cao et al. 2018). Its average detection depth is around 24.5–26 mag (Y. Gong et al. 2019). The large-scale, multiband photometric survey of CSST will offer significant opportunities for identifying a large number of binary star systems. Paper I represents the preliminary work for this project. Based on theoretical simulation data for CSST and a single metallicity assumption, they developed a preliminary classifier using the multilayer perceptron (MLP) framework to distinguish between MS single stars and binary stars. Additionally, they proposed a method for calculating the binary star fraction. However, in their previous work, because they were in the initial stages of methodological design and validation, they only considered a single metallicity, specifically solar metallicity, without accounting for the effects of extinction on photometry. Additionally, photometric errors were simplified to basic Poisson errors. Therefore, in this study, to better align with the actual conditions of future CSST photometric survey data, we carefully evaluated the impact of these factors on the photometric data, reconstructed the simulated data accordingly, and obtained some new results by integrating existing photometric surveys.

In this paper, to make the simulated data more closely resemble the observed data sample, we further enhance our identification model by extending the mock training set to include multiple metallicities, extinction, and photometric errors, and validate it using observed data from Gaia and the Galaxy Evolution Explorer (GALEX). The structure of this article is as follows: In Section 2, we present the simulated data and describe the composition of the observed data used for validation. In Section 3, we give the establishment of the classification model and the derivation of the binary fraction calculation method. In Section 4, we present the results of the simulations and observational validations, while a discussion and summary are provided in Sections 5 and 6.

## 2. Data

To establish the identification model, we first construct mock data with various metallicity, extinction, and photometric errors. We then describe the observational data used for validation.

### 2.1. Mock Data

Similarly to Paper I, the mock data remains focused exclusively on the MS region of the Hertzsprung–Russell diagram, and the detailed steps for generating the mock data are outlined as follows:

(1) *Mock data for single star*. We construct a data grid with masses ranging from 0.1 to 10 $M_\odot$ (with intervals of $\Delta \log M/M_\odot = 0.01$) and metallicities ([Fe/H]) ranging from $-1.3$ to 0.5 (with intervals of $\Delta$[Fe/H] = 0.1). For each specific mass and metallicity value, the evolution track in the MS phase from MIST is extracted and divided into 50 segments of equal spacing in radius, yielding the corresponding stellar parameters, such as effective temperature ($T_{\rm eff}$), surface gravity ($\log g$), and age.

(2) *Mock data for binary*. Based on the single-star sample derived in step (1), we use these single stars as the primary stars in the binary systems. In simulating the binary star sample, we exclude single-star samples with masses less than 0.8 $M_\odot$ from being considered as the primary stars in the binary systems, to ensure the $M_2 > 0.1 M_\odot$ when $q > 0.1$, allowing it to be included in the spectral library. Then we assign a uniformly distributed mass ratio ($f(q)$) to determine the mass of the secondary component ($M_2$). Subsequently, assuming that both the primary and secondary stars share the same evolutionary age, we use interpolation within the track grids, based on their mass and age, to obtain the stellar parameters for the secondary star.

(3) *Apparent magnitudes for mock data in CSST*. To derive the apparent magnitudes for the mock data for single stars

and binaries, we first derive the spectra of stars using their $T_{\text{eff}}$, $\log g$, and [M/H].[7] The BT-Settle spectral library is employed to produce the reference spectra. And the ranges for $T_{\text{eff}}$, $\log g$, and [M/H] in this library are $2600\,\text{K} \leqslant T_{\text{eff}} \leqslant 50{,}000\,\text{K}$, $0.5 \leqslant \log(g/(\text{cm s}^{-2})) \leqslant 6$, $-4.0 \leqslant [\text{M/H}] \leqslant 0.5$. Then, we performed a random matching of the simulated sample with positions (R.A., decl., distance $d$) from the Gaia DR3 all-sky data. By integrating this information with the transmission curve of the CSST photometric system, denoted as $S_{\lambda,i}$, where $i = 1, 2, 3, 4, 5, 6$, and $7$ corresponds to the NUV, $u$, $g$, $r$, $i$, $z$, and $y$ bands, respectively. The apparent magnitudes for a single star ($m_i$) can be calculated by

$$m_i = -2.5 \log \left[ \left(\frac{R}{d}\right)^2 \frac{\int_{\lambda_1}^{\lambda_2} \lambda f_\lambda S_{\lambda,i} d\lambda}{\int_{\lambda_1}^{\lambda_2} \lambda f_\lambda^0 S_{\lambda,i} d\lambda} \right], \quad (1)$$

where $f_\lambda^0$ denotes the flux of the reference spectra, defined by $f_\lambda^0 = \frac{c}{\lambda^2} f_\nu^0$. The value of $f_\nu^0$ is given by $f_\nu^0 = 10^{\frac{48.60}{-2.5}}\,\text{erg} \cdot \text{s}^{-1} \cdot \text{cm}^{-2} \cdot \text{Hz}^{-1}$, as the AB magnitude system is employed in the CSST. The parameters $\lambda_1$ and $\lambda_2$ represent the wavelength limits of each band, while $R$ signifies the stellar radius and $f_\lambda$ refers to the intrinsic flux.

For a binary system, the apparent magnitude ($m_{b,i}$) is determined by

$$m_{b,i} = -2.5 \log \left[ \frac{R_1^2 \int_{\lambda_1}^{\lambda_2} \lambda f_{\lambda 1} S_{\lambda,i} d\lambda + R_2^2 \int_{\lambda_1}^{\lambda_2} \lambda f_{\lambda 2} S_{\lambda,i} d\lambda}{d^2 \int_{\lambda_1}^{\lambda_2} \lambda f_\lambda^0 S_{\lambda,i} d\lambda} \right], \quad (2)$$

where $f_{\lambda 1}$ and $f_{\lambda 2}$ represent the fluxes of the two components in the binaries; $d$ represents the distance of this star from Earth. $R_1$ and $R_2$ correspond to the radius of the primary and secondary, respectively.

(4) *Adding extinction.* Based on the coordinates and distances from the mock data, we used the 3D dust map Bayestar to calculate the extinction (G. M. Green et al. 2019). The extinction curve illustrates the degree of extinction across various wavelengths, commonly denoted by the ratio $R_\lambda$, which quantifies the total extinction relative to the selective extinction $R_\lambda = \frac{A_V}{E(B-V)}$, $R_V = 3.16$. The extinction relative to the $V$ band can be expressed as $f_\lambda = \frac{A_\lambda}{A_V}$. In our CSST extinction simulation, we utilized the value $f_{\text{NUV}} = 2.516$ for GALEX (R. E. Wall et al. 2019). For the $u$, $g$, $r$, $i$, and $z$ bands, we adopted $f_u = 1.584, f_g = 1.205$, $f_r = 0.848, f_i = 0.630$, and $f_z = 0.458$, while for the $y$ band, we used $R_y = 0.395$ (S. Wang & X. Chen 2019). Therefore, the apparent magnitudes $m_i'$ after applying extinction can be calculated by

$$m_i' = m_i + A_\lambda, \quad (3)$$

where $A_\lambda$ represents the extinction value of each band.

(5) *Adding photometric errors.* An overall photometric error of the simulation is considered in this paper, which includes the influence of sky background light $f_{\text{sky}}$, which encompasses both Earth-reflected light $f_{\text{Earth}}$ and zodiacal light $f_{\text{zodiac}}$, along with filter transmission $S_\lambda$, dark current noise $n_{\text{dark}}$, and signal readout noise $n_{\text{readout}}$. With the noise contributions from the aforementioned components, the signal-to-noise ratio (S/N) can be calculated using the following formula:

$$\text{S/N} = \frac{0.8 \times f_{\text{signal}}}{\sqrt{0.8 \times f_{\text{signal}} + f_{\text{sky}} \cdot \text{ex}_{\text{time}} \cdot \text{ex}_{\text{num}} \cdot c_{\text{num}} + n_{\text{dark}} + n_{\text{readout}}}}, \quad (4)$$

where the signal light flux $f_{\text{signal}}$ can be calculated by

$$f_{\text{signal}} = 5.48 \times 10^{10} \cdot 10^{-0.4 \cdot m_i'} \cdot \int \frac{e_f}{\lambda} d\lambda \cdot \text{ex}_{\text{time}} \cdot \text{ex}_{\text{num}} \cdot \pi \left(\frac{\text{aperture}}{2}\right)^2, \quad (5)$$

where $m_i'$ is the given magnitude, $e_f$ is the transmission efficiency of light as it passes through optical devices (such as lenses and filters) at different wavelengths, $\text{ex}_{\text{num}}$ is the number of exposures, and aperture is the size of the telescope aperture. The formula used to calculate the $f_{\text{sky}}$ is presented as follows:

$$f_{\text{sky}} = \int (f_{\text{Earth}} + f_{\text{zodiac}}) \cdot S_\lambda \, d\lambda, \quad (6)$$

where $S_\lambda$ represents the transmission of the instrument at a specific wavelength. $n_{\text{dark}}$ can be derived by

$$n_{\text{dark}} = 0.02 \times \text{ex}_{\text{time}} \cdot \text{ex}_{\text{num}} \cdot c_{\text{num}}, \quad (7)$$

where $\text{ex}_{\text{time}}$ is the time of exposures. $c_{\text{num}} = \pi \cdot r_{\text{pix}}^2$, is a constant related to pixel area, while $r_{\text{pix}}$ is the pixel size calculated based on radius and angular size. $n_{\text{readout}}$ can be derived by

$$n_{\text{readout}} = 25 \times c_{\text{num}} \cdot \text{ex}_{\text{num}}. \quad (8)$$

We conduct a single 150 s exposure based on the CSST survey schedule, so $\text{ex}_{\text{num}} = 1$ and $\text{ex}_{\text{time}} = 150\,\text{s}$. After obtaining the S/N for each star, we can calculate the apparent magnitudes after adding noise $m_i''$ using the following series of formulas:

$$\begin{aligned}
f_{\text{in}} &= 10^{-0.4 \cdot (m_i' + 48.6)}; \\
n_{\text{in}} &= \frac{f_{\text{in}}}{\text{S/N}}; \\
f_{\text{now}} &= f_{\text{in}} + n_{\text{in}}; \\
m_i'' &= -2.5 \cdot \log_{10}(f_{\text{now}}),
\end{aligned} \quad (9)$$

where $f_{\text{in}}$ denotes the flux associated with the input magnitude, $n_{\text{in}}$ represents the corresponding noise, and $f_{\text{now}}$ signifies the flux subsequent to the addition of noise. Based on the above formula, we can derive the relationship between the S/N and magnitude under these simulation conditions. This relationship is illustrated in Figure 1, where we observe that fainter stars exhibit a lower S/N. Additionally, for stars of the same brightness, the S/N is highest in the $g$ band, followed by the $r$, $i$, $z$, and NUV bands in descending order, with the $y$ band showing the lowest S/N.

Following comprehensive discussions with the CSST team, we concluded that the actual observational data in the $g$ band should ideally peak around a magnitude of 22. To align with this requirement, we applied a distance modulus of DM = 10 to the mock data, which enhances the realism of the resulting S/N. Considering that the CSST's bright-end observational threshold is approximately at the 26 mag and recognizing that stars brighter than the 18 mag are likely to be overexposed, we ultimately selected samples with $m_g'' \geqslant 18$ and $m_{\mathrm{NUV}}'' \leqslant 26$ for our final simulation data set, which comprises a total of 656,326 systems. Figure 2 illustrates the magnitude distribution of the mock data for single stars and binaries across the seven bands of the CSST with colored histograms.

---

[7] The [Fe/H] values obtained from MIST are converted to [M/H] using the formula $[\text{M/H}] = [\text{Fe/H}] + \log 10(0.638f + 0.362)$, where $f = 1$ for $[\text{Fe/H}] \leqslant 0.5$, and $f = 2.91$ for $[\text{Fe/H}] < 0.5$ (M. Salaris et al. 1993).

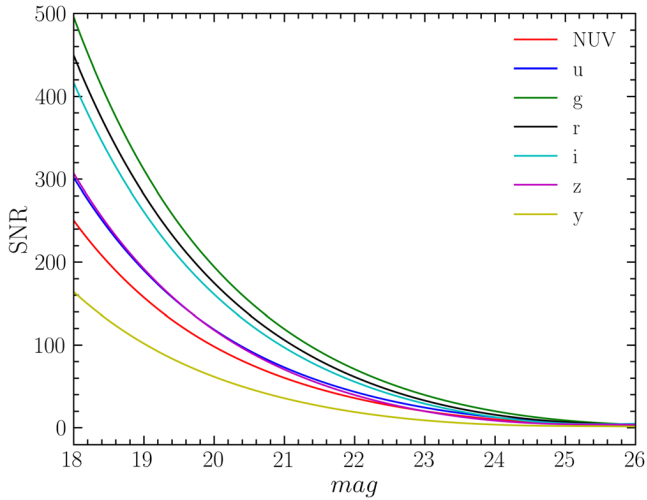

**Figure 1.** The relationship between the S/N of CSST and the apparent magnitudes across seven bands as denoted in the plot.

### 2.2. Observed Data

#### 2.2.1. Sample Selection

To validate the model, we selected data from existing photometric surveys in various bands and compared them with the seven bands of the CSST. We first categorized the stars into single and binary systems based on Gaia DR3 observations (Gaia Collaboration et al. 2023a). Subsequently, we obtained photometric data in seven bands from (Gaia Collaboration et al. 2023a), SDSS (P. Montegriffo et al. 2023), Pan-STARRS1 (M. Fouesneau et al. 2023), and GALEX (L. Bianchi et al. 2014).

For the single-star observational sample, we selected the OBA-, FGKM-, and solar-type stars from Gaia's golden sample[8,9,10] (Gaia Collaboration et al. 2023b) as the initial data set. These stars excluded all sources identified as variable or non-single stars (NSSs) based on Gaia's astrometric data, and were observed with high-quality astrophysical parameters from Gaia, including effective temperatures and surface gravities. First, we performed quality control on the initial data, retaining targets with RUWE < 2.0 and parallax_over_error > 5. Then, we excluded the targets that are not in the MS phase.

Although Gaia's golden sample has already excluded variable and NSSs identified by Gaia, other time-domain photometric surveys such as Kepler, TESS, Zwicky Transient Facility (ZTF), and ASAS-SN, which are also observed by Gaia, have identified hundreds of thousands of variable and binary stars based on their light curves. Therefore, to ensure that any potential variable or binary stars, not detected by Gaia but identified by these other surveys, additional exclusion criteria based on these light curves were applied. To achieve this, we crossmatched these stars with Gaia's NSS catalogs[11,12,13,14,15] and excluded the known eclipsing binaries identified in ASAS-SN (D. M. Rowan et al. 2022, 2023a, 2023b), ZTF (X. Chen et al. 2020), TESS (A. Prša et al. 2022; L. W. IJspeert et al. 2021), and Kepler (R. W. Slawson et al. 2011; A. Prša et al. 2011).

---

[8] gaiadr3.gold_sample_oba_stars
[9] gaiadr3.gold_sample_solar_analogues
[10] gaiadr3.gold_sample_fgkm_stars

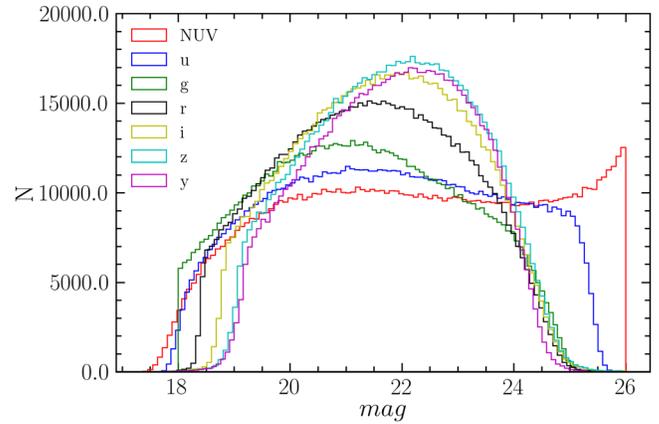

**Figure 2.** Histogram of the apparent magnitude distribution for the mock samples. Different colors correspond to the seven bands (NUV, u, g, r, i, z, y) of CSST.

For the binary star sample, our initial data were entirely sourced from Gaia's NSS catalogs. Similarly, we retained only those sources with RUWE < 2.0 and parallax_over_error > 5. Finally, we obtained a total of 415,863 single-star samples and 134,491 binary star samples from the observations.

#### 2.2.2. Photometric Data Collection

To obtain the photometric data in the NUV, u, g, r, i, z, and y bands, we used the Gaia Synthetic Photometry Catalogue (GSPC). The GSPC provides the u-, g-, r-, i-, z-, and y-band data based on the standardized SDSS photometric system. We crossmatched the compiled single-star and binary samples with the GSPC to retrieve their photometric data in these bands. Additionally, we crossmatched these samples with the GALEX survey to obtain the NUV photometric data. A crossmatch radius of 2″ was applied. To align the observational data with the future CSST detection scenario, we increased the observed magnitudes by 10 mag to bring the peak close to around 22 mag. Figure 3 illustrates the apparent magnitude distribution in the seven bands of the single-star and binary samples compiled with colored histograms.

### 3. Method

#### 3.1. Model Establishment

This section will introduce our binary and single-star classification model based on the MLP in Section 3.1.1, followed by the development of a method for calculating the binary star ratio based on the results of this model in Section 3.1.2.

---

[11] gaiadr3.nss_acceleration_astro
[12] gaiadr3.nss_non_linear_spectro
[13] gaiadr3.nss_two_body_orbit
[14] gaiadr3.nss_vim_fl
[15] gaiadr3.binary_masses

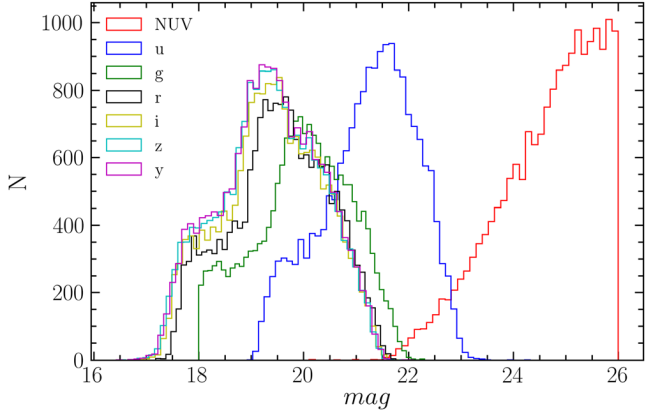

**Figure 3.** Histogram of the apparent magnitude distribution for the observed samples. Different colors correspond to the seven bands (NUV, *u*, *g*, *r*, *i*, *z*, *y*).

*3.1.1. Binary Identification*

We have constructed a mock sample comprising a total of 656,326 systems. Of these, 500,000 systems were designated for the training set, with the remaining samples allocated for testing. The training set includes 250,000 single stars and 250,000 binary stars. In this study, we employed the MLP model to distinguish between single stars and binary stars in a mixed sample. The MLP classifier model is a powerful supervised learning algorithm widely used for classification tasks. The inputs of our MLP model are the apparent magnitudes from seven bands (NUV, *u*, *g*, *r*, *i*, *z*, and *y* bands) for the mock data, the outputs are 0 or 1, where 0 indicates a classification as a single star and 1 indicates a classification as a binary star. The model comprises three hidden layers with 70, 70, and 30 neurons, respectively.

To introduce nonlinearity into the model, we utilized the rectified linear unit activation function. The model is trained using the Adam optimizer, which adaptively adjusts the learning rate to facilitate faster convergence. We set the initial learning rate of 1e-4 and incorporated a momentum term of 0.9 to accelerate the movement of gradient vectors in the correct direction, thereby promoting quicker convergence. The classifier also employs an early stopping mechanism that monitors validation loss and halts training if there is no improvement over 50 iterations. Additionally, we used a batch size of 10,000 to efficiently process the data, while the tolerance for stopping criteria is set to 1e-5. This architecture and parameter configuration demonstrated faster training speeds and higher accuracy in our extensive testing. After training, we employed the F1 score to evaluate the performance of our model, the F1 score, calculated as the harmonic mean of precision and recall, reflects the model's performance based on the confusion matrix. And the F1 score is defined as

$$F1 = \frac{2 \times P \times R}{P + R}, \quad (10)$$

where $P$ denotes the proportion of true binaries among all systems classified as binaries, while $R$ represents the fraction of true binaries identified out of the total binaries in the mock sample. A higher F1 score suggests a more reliable model.

For each system in the test sample, whether it is single or binary, we use the trained MLP model to calculate the probability of it being a binary, $p_b$, or a single star, $p_s$. If $p_b \geqslant 0.5$ (or $p_s \geqslant 0.5$), the system will ultimately be classified as a binary (or single star). The classification results are then organized into four categories, as shown in Figure 4.

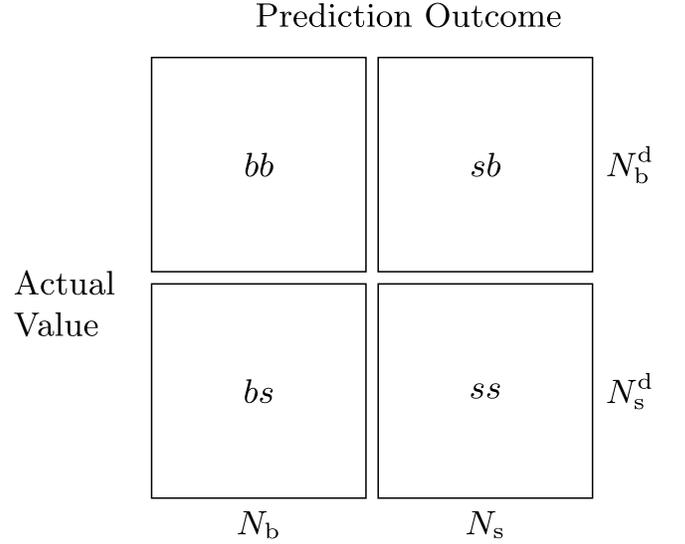

**Figure 4.** Confusion matrix: bb represents the number of binaries detected as binaries in our MLP models; sb represents the number of single stars detected as binaries in our MLP models; and bs represents the number of binaries detected as single stars in our MLP models. ss represents the number of single stars detected as single stars in our MLP models. This figure is consistent with Figure 3 in Paper I, and it is included here to provide a clearer explanation of the methodology.

In Figure 4, "bb" represents true binaries detected as binaries, "bs" represents true binaries detected as single stars, "ss" represents true single stars detected as single stars, and "sb" represents true single stars detected as binaries. $N_{bb}$, $N_{sb}$, $N_{bs}$, and $N_{ss}$ represent the number of systems in each category, the total number of binaries and single stars in the mock sample can be expressed as $N_b = N_{bb} + N_{bs}$ and $N_s = N_{sb} + N_{ss}$, respectively. The precision $P$ is then given by $N_{bb}/(N_{bb} + N_{sb})$, and the recall $R$ by $N_{bb}/N_b$. For each trained MLP model, we evaluate its performance using a binary detection efficiency, $\rho_b = N_{bb}/N_b$, and a single star detection efficiency, $\rho_s = N_{ss}/N_s$. These metrics reflect the model's effectiveness in identifying binaries and single stars, respectively.

*3.1.2. Binary Fraction*

In a sample consisting of $N$ systems, where $N_b$ represents the number of binaries and $N_s$ indicates the number of single stars, the number of detected binaries, denoted as $N_b^d$, can be computed using the following equations after applying our trained MLP models. The true binary sample size $N_b$, presented in Paper I, can be expressed as

$$N_b = \frac{(1 - \rho_s)N - N_b^d}{1 - \rho_b - \rho_s}. \quad (11)$$

The corresponding binary fraction in the sample is

$$f_b = \frac{1 - \rho_s - f_b^d}{1 - \rho_b - \rho_s}, \quad (12)$$

where $f_b^d = N_b^d/N$ is the detected binary fraction of the sample. Using Equations 11 and 12, we can apply the concept of calculus to approximate the true number of binaries and their proportion by integrating the detection efficiency of the model at each small $q$ interval.

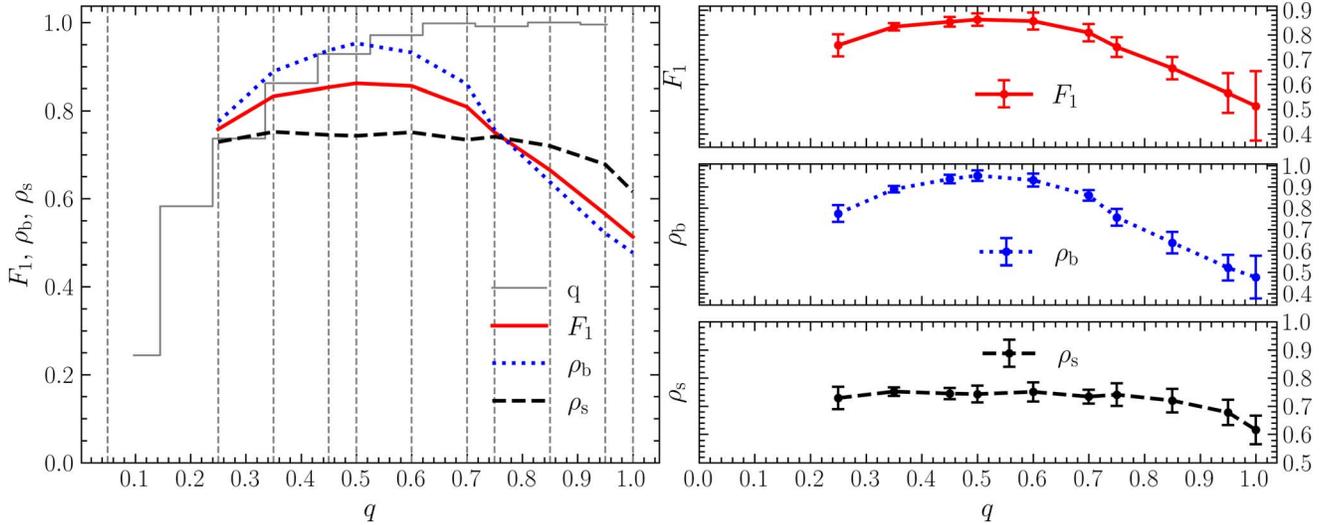

**Figure 5.** Dependence of F1 (red solid line), $\rho_b$ (blue dotted line), and $\rho_s$ (black dashed line) on the mass ratio. The solid gray line in the left panel represents the normalized histogram of the mock sample mass ratio. The gray dashed vertical lines indicate the boundaries of the bin widths we sampled. In the right panel, the points represent the mean from 1000 resampling tests for each data point, and the error bars denote the corresponding standard error. The error bars in different colors correspond to different variables: red represents F1, blue represents $\rho_b$, and black represents $\rho_s$.

### 3.2. Model Validation

#### 3.2.1. Detected Efficiency in Mass Ratio

The testing set from the mock data is used to validate our model. Figure 5 illustrates the dependence of F1, $\rho_b$, and $\rho_s$ on the mass ratio $q$ in our MLP model. In the left panel of Figure 5, the solid gray histogram shows the distribution of the testing set, including 134,491 binaries and 415,863 single stars. To address statistical errors caused by uneven sample distributions, we then divided the samples into 10 subsamples across $q$, ensuring equal sample sizes in each bin. The range of $q$ for each bin is indicated by the vertical gray dashed lines in the left panel. The performance of F1, $\rho_b$, and $\rho_s$ is shown in the red solid line, blue dotted, and black dashed lines, respectively. Additionally, we randomly sampled 50% of the test sample for 1000 resampling iterations to reevaluate F1, $\rho_b$, and $\rho_s$. The mean values from these 1000 samples are marked by points in the right panel of Figure 5, and the standard deviations are represented by error bars (bootstrap method).

As shown in Figure 5, the model exhibits a lower detection rate for binaries at extreme mass ratios, while achieving over 80% detection efficiency for binaries with $q$ between 0.3 and 0.7. The model has learned the physically based result that binary systems with extreme mass ratios are more difficult to detect: when $q \approx 0$, the companion's contribution is negligible, making it easier to classify the system as a single star, whereas when $q \approx 1$, the colors of the two-member stars in a binary system are challenging to distinguish. We can observe that when $q > 0.7$, not only does the F1 decrease but the detection efficiency for binaries $\rho_b$ also declines, with a significant increase in the corresponding error. This indicates that the model is finding it increasingly challenging to classify samples in this range. In contrast, the detection efficiency for single stars remains approximately constant across different mass ratios. The absence of a notable change in detection efficiency for single stars with respect to $q$ suggests that our assumption of assigning a random $q$ value to single stars is valid in this context.

#### 3.2.2. Detected Efficiency in $m_g$

In addition, we applied the same method to investigate the dependence of the $g$-band apparent magnitude $m_g$ on F1, $\rho_b$, and $\rho_s$. From Figure 6, we can see that for magnitudes ranging from 18 to 24, as the apparent magnitude increases and the stars become fainter, the model's detection efficiency for binaries $\rho_b$ shows a noticeable decline starting at 0.7, with increasing error bars. This indicates that detecting fainter stars poses a greater challenge for our model. In the last sampling bin, where $m_g$ is approximately between 24 and 26, F1 and $\rho_s$ exhibit an upward trend. However, the corresponding error bars are very large, suggesting that this trend is likely due to a combination of the decreased detection efficiency for faint stars and the small sample size in that bin, resulting in statistical errors.

#### 3.2.3. Validation of the $f_b$ Calculation Method in Mock Data

If we apply the MLP model directly to a batch of samples to obtain the detected number of binaries $N_b^d$, we can observe from the leftmost panel of Figure 7 that it does not correspond well to the true number of binaries in the sample. In this case, we utilize the binary number calculation method described in Section 3.1.2 to recompute the number of binaries in the sample according to Equation (11), incorporating the model results presented in Figure 5.

We constructed 20 test samples, each containing a fixed number of stars $N = 100,000$. For the number of binaries $N_b^d$ in each sample, we varied this value from 5000 to $N$–5000 in increments of 5000. The corresponding binary fraction $f_b$ ranged from 5% to 95%, with a step of 5%. For each test sample, we assumed the mass ratio distribution to be one of the four types illustrated in Figure 8: uniform, normal, exponential, and negative exponential distributions. We used this assumption as input for Equation (11) to validate the binary fraction calculation method.

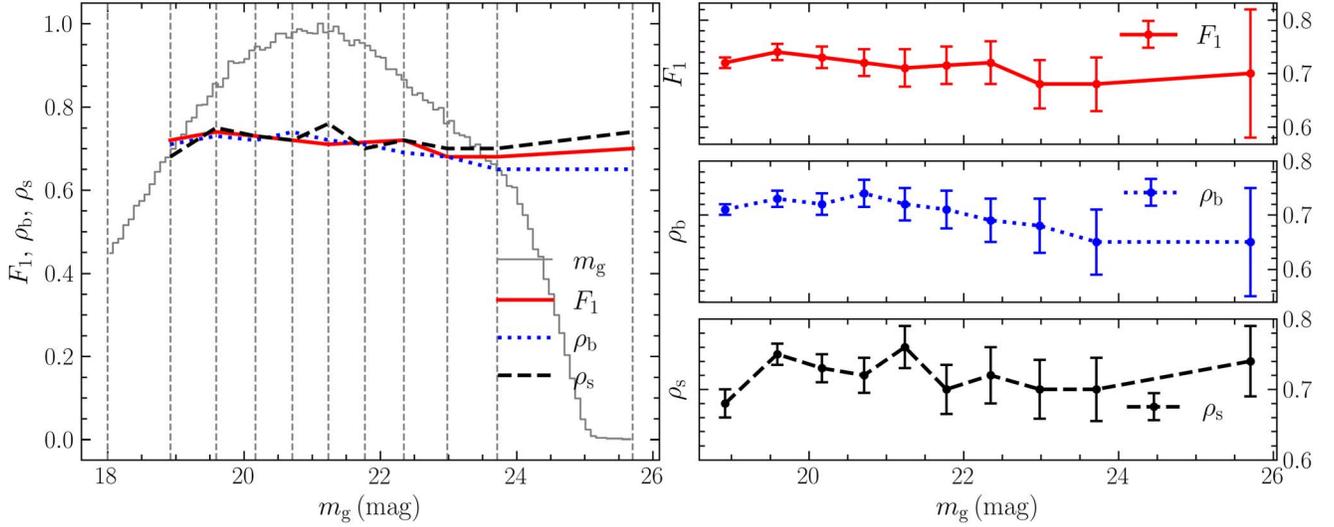

**Figure 6.** The relationship of F1 (red solid line), $\rho_b$ (blue dotted line), and $\rho_s$ (black dashed line) on the $m_g$ for the mock sample. The solid gray line in the left panel represents the normalized histogram of the mock sample $m_g$. The gray dashed vertical lines indicate the boundaries of the bin widths we sampled. In the right panel, the points represent the mean from 1000 resampling tests for each data point, and the error bars denote the corresponding standard error. The error bars in different colors correspond to different variables: red represents F1, blue represents $\rho_b$, and black represents $\rho_s$.

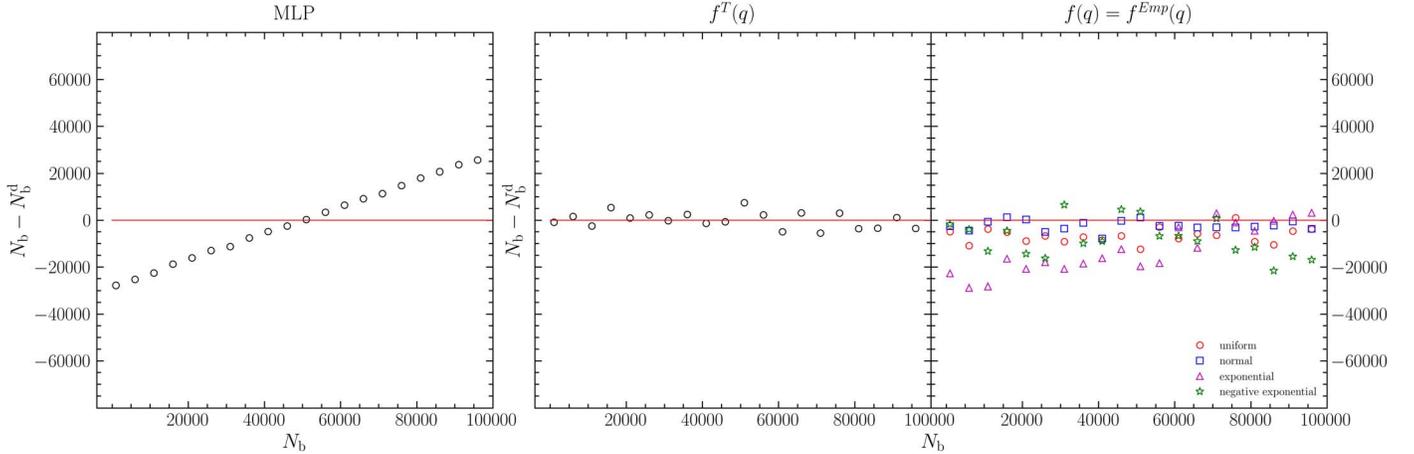

**Figure 7.** The relationship between the true number of binaries $N_b$ and the inferred number of binaries $N_b^d$ is examined in the mock sample. In the left column, the inferred number of binaries corresponds to the count obtained from our trained MLP model. The middle panel employs the exact same mass ratio distribution as the test sample to estimate the number of binaries, as described in Section 3.1.2. In the right column, in the absence of mass ratio distribution information, we assume a uniform mass ratio distribution to infer the number of binaries using the method outlined in Section 3.1.2. The symbols—open circle (red), open square (blue), filled triangle (purple), and open star (green)—represent uniform, normal, exponential, and negative exponential mass ratio distributions of the test samples, respectively.

Subsequently, we tested these constructed samples using the trained MLP model to obtain $N_b^d$. The differences between $N_b^d$ and $N_b$, as shown in the left column of Figure 7, reveal that the overall detected number of binaries $N_b^d$ deviates significantly from the true binary number $N_b$ in the sample due to the model's lower detection rate at extreme mass ratios. However, when we input the true mass ratio distribution $f^T(q)$ for this test sample and utilize Equation (11) for binary number calculations, the detected number of binaries $N_b^d$ becomes very close to $N_b$, as illustrated in the middle column of Figure 7.

The mass ratio distribution of the simulated sample is uniform. However, the mass ratio of the test sample is unknown when calculating the number of binary stars. Therefore, we assume that its distribution follows one of the four types depicted in Figure 8, we find that when the assumed mass ratio distribution is normal, $N_b^d$ closely approximates $N_b$, likely due to the model's high detection rate within the range $0.3 < q < 0.7$. When the assumed mass ratio distribution is uniform, it is very close to the actual sample mass ratio distribution $f^T(q)$, resulting in $N_b$ and $N_b^d$ being similarly close. However, the detection of binaries $N_b^d$ remains lower than the true binary number $N_b$ due to the model's poor detection capability at extreme mass ratios. In contrast, when the assumed mass ratio distribution is exponential or negative exponential, the deviation from the true mass ratio $f^T(q)$ is substantial, as shown in the rightmost column of Figure 7. Nonetheless, even when the assumed sample mass ratio distribution differs from the true distribution, $N_b^d$ remains close to $N_b$, validating the effectiveness of our binary number calculation method.

## 4. Results

After validating the model on the mock data, we now apply it to real observational data for further evaluation. The observational sample consists of 2761 binary sources with mass ratio determinations and 415,863 single sources obtained through the sample selection method described in Section 2.2. Using the same testing method as in Section 3.2, we divided the test sample into 10 small bins based on mass ratio, ensuring equal sample sizes in each bin, as shown in Figure 9. Due to the limited number of sources with mass ratio solutions among binaries after crossmatching with GALEX, we performed data augmentation to reduce statistical errors. We performed a two-fold proportional expansion of the sample. In Figure 9, the blue line represents the mass ratio distribution of binary stars in the sample, with a total of 2761 binaries. It can be observed that certain bins, such as $0.6 < q < 0.7$. To maintain a distribution profile closer to the original sample and to achieve smoother results, we randomly extracted additional samples to bring the total number in this bin to 500. Then, we constructed an expanded test sample $N = 10{,}000$ that includes all binary samples with $N_b = 5000$, and a randomly selected single-star sample of the same size, $N_s = 5000$. The gray line shows the distribution after extending the sample, with a total of 5000 binaries. The distribution of the expanded sample is presented in Figure 9. Finally, we tested the responses of F1, $\rho_b$, and $\rho_s$ in relation to $q$ for each bin in the test sample, yielding the results shown in Figure 9

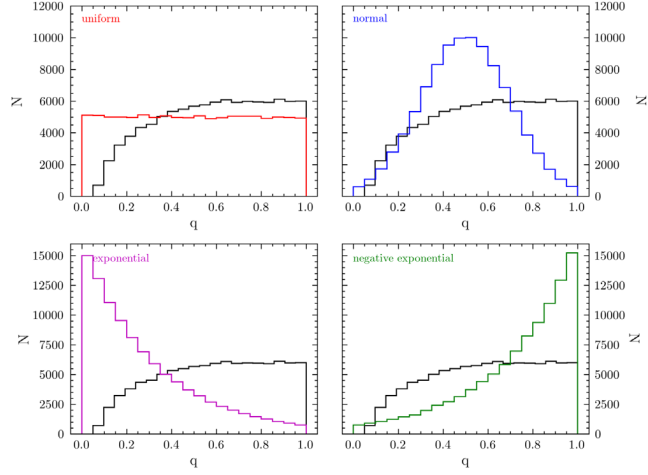

**Figure 8.** The mass ratio distribution of the mock sample (black) and the four empirical mass ratio distributions used as hypothesized inputs: uniform (red), normal (blue), exponential (purple), and reverse exponential (green).

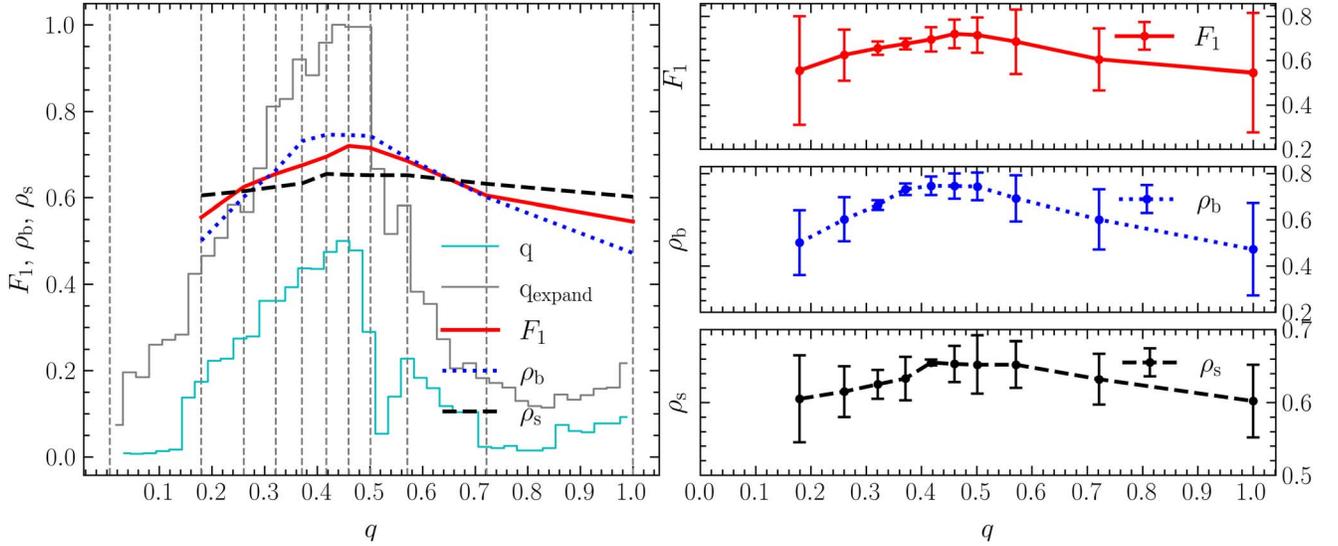

**Figure 9.** Dependence of F1 (red solid line), $\rho_b$ (blue dotted line), and $\rho_s$ (black dashed line) on the mass ratio. The solid cyan line in the left panel represents the normalized histogram of the observed sample's mass ratio with the NUV band. The solid gray line in the left panel represents the normalized histogram of the expanded observed sample's mass ratio with the NUV band. The gray dashed vertical lines indicate the boundaries of the bin widths we sampled. In the right panel, the points represent the mean from 1000 resampling tests for each data point, and the error bars denote the corresponding standard error. The error bars in different colors correspond to different variables: red represents F1, blue represents $\rho_b$, and black represents $\rho_s$.

### 4.1. Detected Efficiency in Mass Ratio

From Figure 9, we can see that the model's performance on the observational samples is slightly inferior to that presented in Figure 5, which is based on simulated samples. For the mass ratio range $0.3 < q < 0.7$, the binary detection rate varies between 68% and 75%. This is likely due to the non-purity of single stars within the observational sample. Some sources labeled as single stars are not actually single, and misclassifications of these stars have led to a lower overall detection rate compared to the simulated data. However, it is worth noting that the model still demonstrates reliable stability when applied to observational data. Similar to the validation results for simulated data, identifying binaries with extreme mass ratios proves challenging, while the detection rate for binaries with intermediate mass ratios averages above 70%. The error increases as $q$ approaches 1.

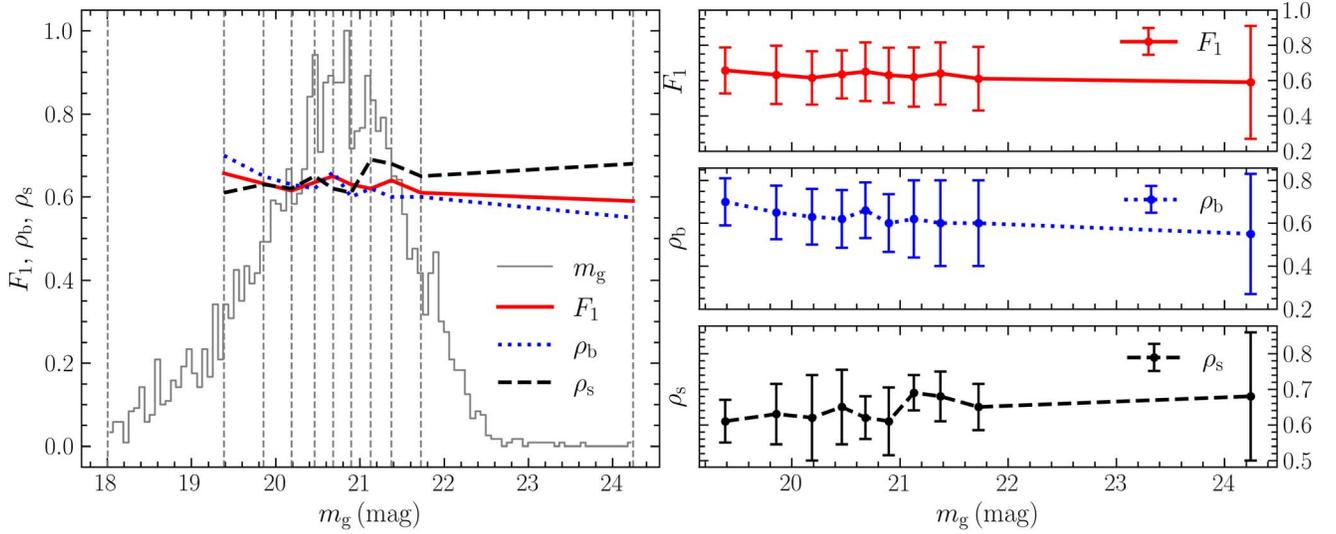

**Figure 10.** Dependence of F1 (red solid line), $\rho_b$ (blue dotted line), and $\rho_s$ (black dashed line) on the $m_g$ for observed sample. The solid gray line in the left panel represents the normalized histogram of the observed sample $m_g$. The gray dashed vertical lines indicate the boundaries of the bin widths we sampled. In the right panel, the points represent the mean from 1000 resampling tests for each data point, and the error bars denote the corresponding standard error. The error bars in different colors correspond to different variables: red represents F1, blue represents $\rho_b$, and black represents $\rho_s$.

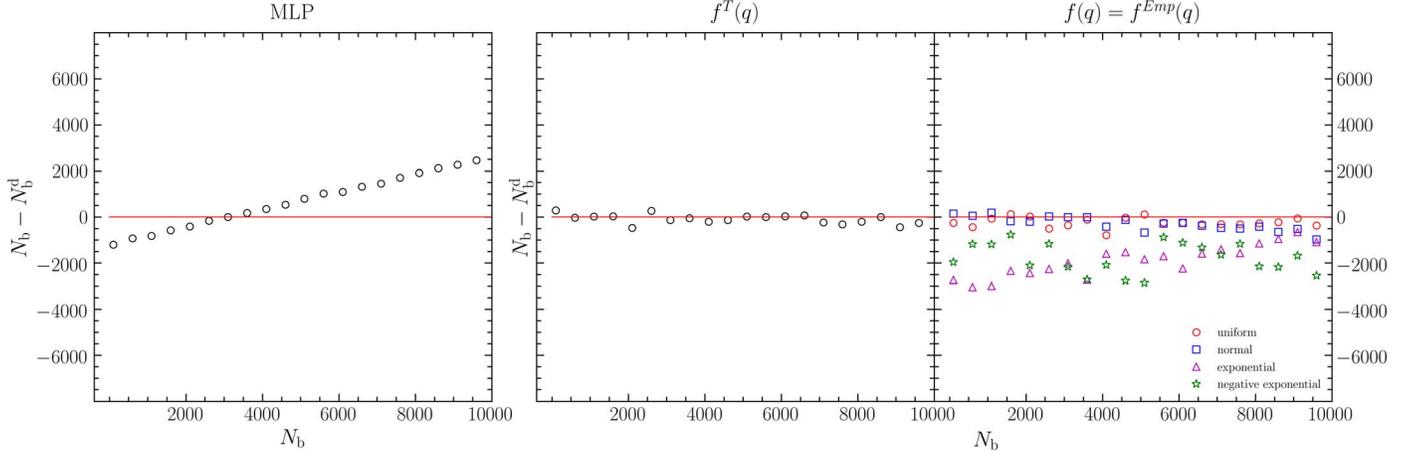

**Figure 11.** The relationship between the true number of binaries $N_b$ and the inferred number of binaries $N_b^d$ is examined in the observed sample with the NUV band. In the left column, the inferred number of binaries corresponds to the count obtained from our trained MLP model. The middle panel employs the exact same mass ratio distribution as the test sample to estimate the number of binaries, as described in Section 3.1.2. In the right column, in the absence of mass ratio distribution information, we assume a uniform mass ratio distribution to infer the number of binaries using the method outlined in Section 3.1.2. The symbols—open circle (red), open square (blue), open triangle (purple), and open star (green)—represent uniform, normal, exponential, and negative exponential mass ratio distributions of the test samples, respectively.

### 4.2. Detected Efficiency in $m_g$

Figure 10 shows the detection efficiency (F1, $\rho_b$, and $\rho_s$) as a function of the $g$-band apparent magnitude, $m_g$. An overall F1, $\rho_b$, and $\rho_s$ is shown in the left panel. In the left panel, the $g$-band apparent magnitude distribution is shown as a gray histogram, while the red solid line, blue dotted line, and black dashed line represent the results for F1, $\rho_b$, and $\rho_s$. In the right panel, the uncertainties of F1, $\rho_b$, and $\rho_s$ are calculated using a bootstrapping algorithm. From Figure 10, it is evident that the model demonstrates a clear trend: as the apparent magnitude increases, indicating fainter stars, the detection rate of binaries decreases. Throughout the entire sample range, an increase of 5 mag in apparent magnitude corresponds to a decrease of approximately 20% in $\rho_b$. Additionally, the right column of Figure 10 shows that as stars become fainter, the resampling error bar increases. At this stage, the model's uncertainty dominates, leading to slightly higher detection efficiency for fainter stars.

### 4.3. Validation of the $f_b$ Calculation Method in Observed Data

Similar to the validation method in Section 3.2.3, we reconstructed 20 test samples controlling for $N = 5000$, with $N_b$ ranging from 250 to 4500 in increments of 250. We obtained results similar to those presented in Figure 11. In the left column of Figure 11, it is shown that the number of binaries detected directly by the MLP model exhibits significant errors when the binary fraction is either very low or very high. However, when we recalculated the number of binaries based on the binary fraction calculation method described in Section 3.1.2, we observed in the middle column of Figure 11 that when the assumed mass ratio distribution aligns with the true mass ratio distribution of the actual sample, the values of $N_b$ and $N_b^d$ exhibit minimal differences. When the mass ratio distribution of the test sample is unknown, and we assume it to be uniform, normal, exponential, or negative exponential as shown in Figure 12, we find that when the assumed mass ratio distribution is uniform and close to the true distribution of the test sample, $N_b^d$ is very close to $N_b$. Furthermore, when the assumed mass ratio distribution is normal, the MLP model shows higher detection efficiency for intermediate mass ratios, resulting in $N_b^d$ also being close to $N_b$. Conversely, when the assumed mass ratio distribution deviates significantly from the true distribution, $N_b^d$ also diverges noticeably from $N_b$. Thus, the test results based on the actual sample are consistent with those from the simulated data, indicating that our model is highly stable and demonstrates reliable performance on observed samples.

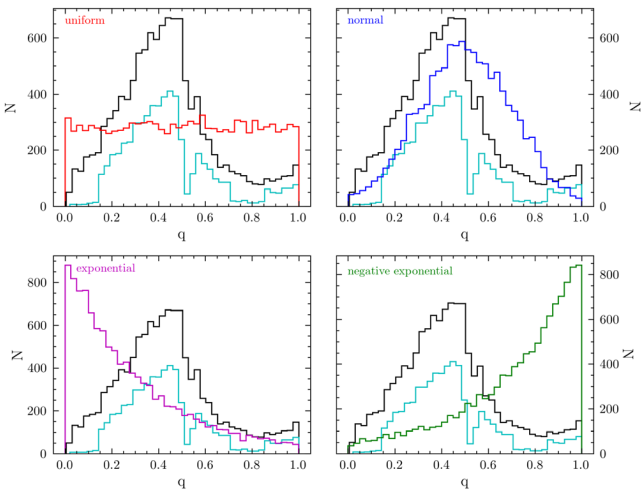

**Figure 12.** The mass ratio distribution of the observed sample with the NUV band (cyan), the mass ratio distribution of the expanded observed sample with the NUV band (black) and the four empirical mass ratio distributions used as hypothesized inputs: uniform (red), normal (blue), exponential (purple), and reverse exponential (green).

## 5. Discussion

Due to the limitations posed by the small number of samples obtained through crossmatching in observational data sets, as well as the fact that some data from certain bands may not be immediately available after future CSST operations, it is necessary to validate whether our method can effectively operate in the absence of certain photometric data. Therefore, this discussion will focus on the implications of lacking NUV-band data.

### 5.1. Model Validation in Mock Data without the NUV Band

Following the same construction method for the training and test sets described in Section 3.1, we removed the NUV-band magnitudes from the input of the model, which now consists of the magnitudes from the $u$, $g$, $r$, $i$, $z$, and $y$ bands. The distribution of these magnitudes is shown in Figure 2. We retrained the MLP classification model in the absence of NUV photometric data and tested it using the same methodology as in Section 3.2, obtaining the F1 score and the responses for $\rho_b$ and $\rho_s$ as a function of $q$, which are presented in Figure 13.

For the mock data, by comparing Figures 5 and 13 for the seven-band and six-band model inputs, we observe a decrease in model performance due to the lack of NUV information; specifically, the binary detection rate $\rho_b$ at $q = 0.5$ dropped from 95% to 80%. This decline is not surprising, as we expected the model's classification performance to diminish with fewer inputs. However, even with the absence of NUV-band input, the model maintains a binary detection rate of nearly 70% for $q < 0.7$, indicating that it still retains reliable performance. Moreover, we note that in Figure 13, the model's errors for $q > 0.7$ are somewhat larger when NUV is excluded, suggesting that the NUV data plays a significant role in the model's accuracy. In our previous work, we also tested the impact of removing other bands on the model, as shown in Figures 12 and 13 of Paper I). The performance of the model is not strongly dependent on the number of input magnitudes. Due to the scheduling of the telescope's survey, the NUV band of CSST will actually be acquired later than other bands. In addition, in this study, the simulated data included factors such as metallicity and extinction, so we repeated this test to ensure that we can continue our work in the early stages of the telescope's operation.

### 5.2. Model Validation in Observed Data without the NUV Band

Using the sample construction method described in Section 2.2, we obtained a binary sample with mass ratio solutions that is not cross-referenced with GALEX, specifically one that does not include NUV-band data. This resulted in 134,491 binary sources and 415,863 single-star sources. We mixed and shuffled the binary and single-star samples, utilizing the trained model that excludes NUV input. Following the principle of having an equal number of stars in each small bin, we divided the samples into 10 bins for testing. We also randomly selected 50% of each test sample for 1000 tests, resulting in the data shown in Figure 14.

Comparing Figure 14 with Figure 13, we observe that the model trained on simulated data shows a decrease of about 10% in F1, $\rho_b$, and $\rho_s$ when validating the real observational samples. However, the test results from the real samples maintain a certain stability in the curves of F1, $\rho_b$, and $\rho_s$ compared to the simulated samples. For instance, the model also demonstrates lower detection efficiency at extreme mass ratios during the validation of the observational samples. At $q = 0.5$, $\rho_b$ still reaches around 0.7, with an overall binary detection rate of 65%.

In contrast, the errors displayed in the right panel of Figure 14 are larger than those in Figure 13. We believe this may be due to the lack of purity in the observational single-star samples. In other words, the overall performance of the observational validation results being worse than that of the simulated data is likely attributable to the fact that sources labeled as single stars in the test sample are not actually single stars.

### 5.3. Validation of the $f_b$ Calculation Method in Observed Data without the NUV Band

Following the testing method described in Section 3.2, and in order to maintain consistency with the simulated data, we constructed 20 test samples from the observational samples described in Section 5.1, with a total number of stars $N = 100,000$, while keeping $N_b$ ranging from 5000 to $N$–5000 in increments of 5000. In Figure 16, we present the differences between the detected number of binaries $N_b^d$ and the true number of binaries $N_b$ for three scenarios: directly testing using the MLP model, inputting the true mass ratio distribution, and inputting four different empirical mass ratio distributions that we hypothesize as shown in Figure 15.

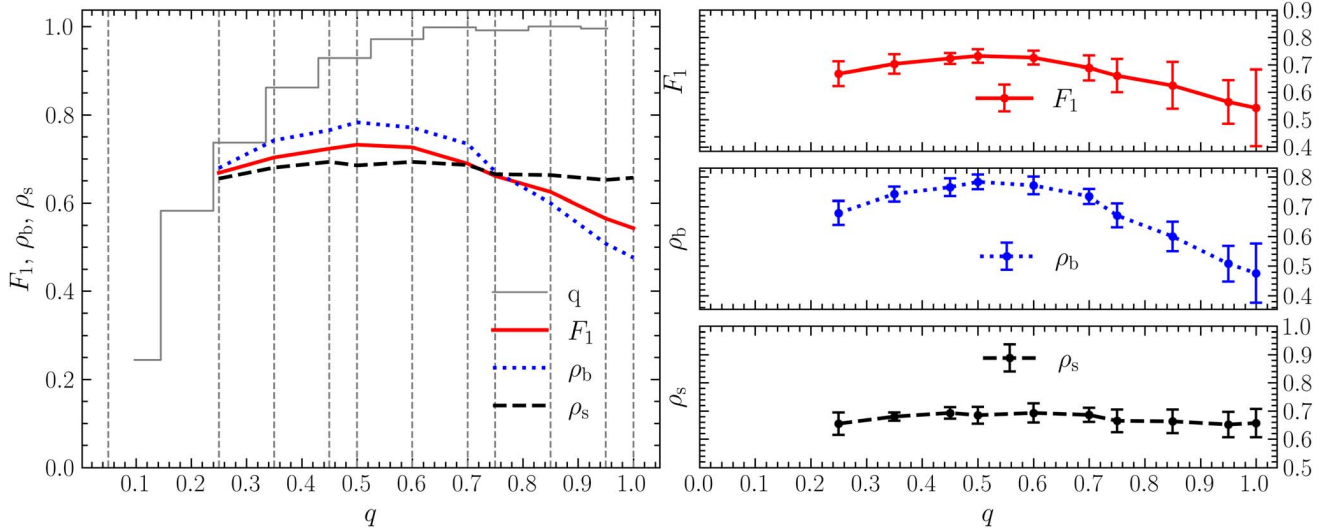

**Figure 13.** The relationship of F1 (red solid line), $\rho_b$ (blue dotted line), and $\rho_s$ (black dashed line) on the $m_g$ for mock sample without the NUV band. The solid gray line in the left panel represents the normalized histogram of the mock sample $m_g$. The gray dashed vertical lines indicate the boundaries of the bin widths we sampled. In the right panel, the points represent the mean from 1000 resampling tests for each data point, and the error bars denote the corresponding standard error. The error bars in different colors correspond to different variables: red represents F1, blue represents $\rho_b$, and black represents $\rho_s$.

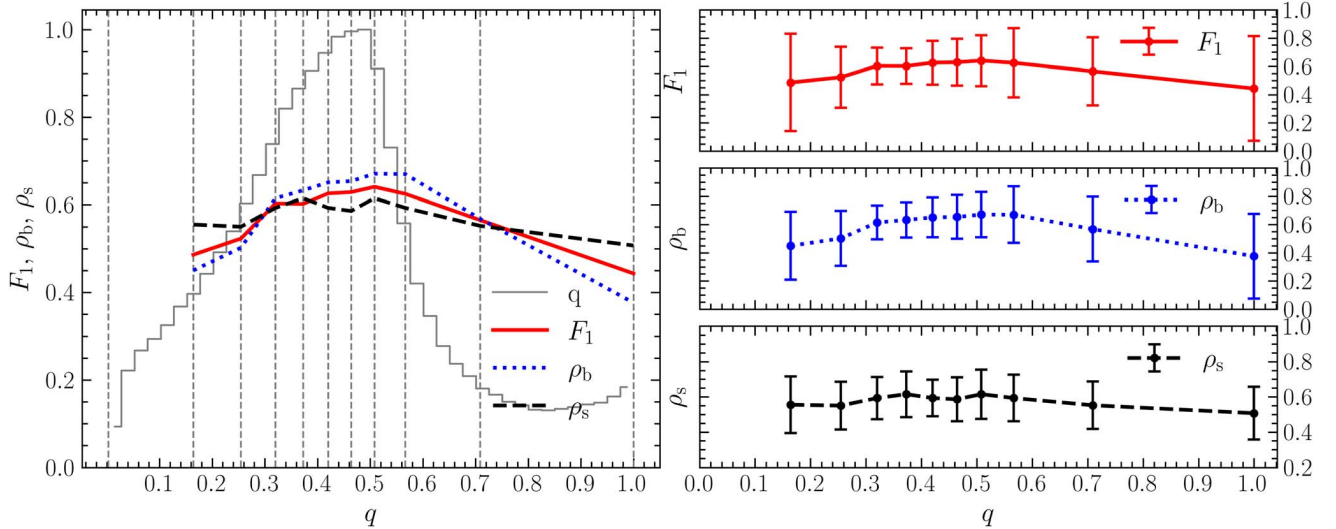

**Figure 14.** The relationship of F1 (red solid line), $\rho_b$ (blue dotted line), and $\rho_s$ (black dashed line) on the $m_g$ for observed sample without the NUV band. The solid gray line in the left panel represents the normalized histogram of the mock sample $m_g$. The gray dashed vertical lines indicate the boundaries of the bin widths we sampled. In the right panel, the points represent the mean from 1000 resampling tests for each data point, and the error bars denote the corresponding standard error. The error bars in different colors correspond to different variables: red represents F1, blue represents $\rho_b$, and black represents $\rho_s$.

It is evident that this method for calculating the number of binaries maintains a high degree of reliability when applied to observational samples. When the true mass ratio distribution $f^T(q)$ is input, the difference between $N_b$ and $N_b^d$ is minimal. When the input is a normal distribution of mass ratios, the difference remains extremely small due to the close resemblance to the distribution of the observational sample.

In contrast, when inputting a uniform distribution $f(q)$, the actual sample is concentrated in the range $0.3 < q < 0.6$, where the model's detection efficiency is higher, leading to a greater number of correctly classified binaries; thus, the difference between $N_b$ and $N_b^d$ is relatively minor as shown in the right column of Figure 16. However, when the input $f(q)$ follows an exponential or negative exponential distribution, the hypothesized mass ratio distribution significantly deviates from the actual sample distribution, resulting in a larger discrepancy between $N_b$ and $N_b^d$.

## 6. Summary

In this study, to distinguish binaries from the upcoming CSST survey, we developed an MS binary detection model based on MLP, incorporating factors such as metallicity, extinction, and photometric errors. The detection model is built using the mock data. The inputs to the model are the apparent magnitudes in the NUV, $u$, $g$, $r$, $i$, $z$, and $y$ bands. The model is then validated using both the mock data and the compiled observational data from Gaia and GALEX. The results illustrate that our method is reliable and efficient in identifying binaries with mass ratios between 0.3 and 0.7, and it shows a significant improvement in the number of binaries identified in the observed sample. After excluding the NUV band, the model's detection efficiency decreases.

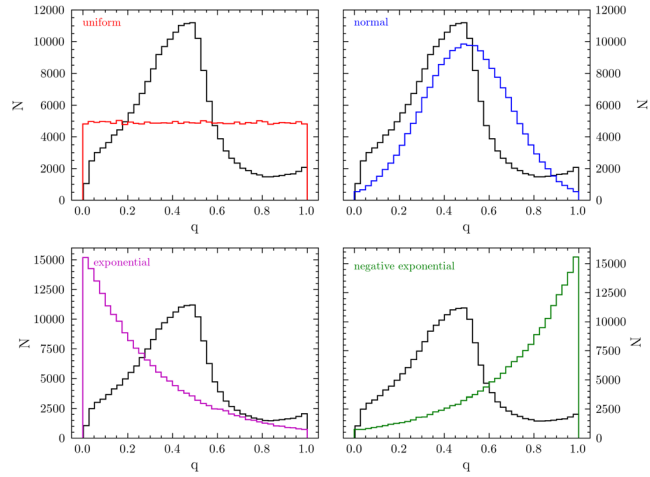

**Figure 15.** The mass ratio distribution of the observed sample without the NUV band (black) and the four empirical mass ratio distributions used as hypothesized inputs: uniform (red), normal (blue), exponential (purple), and reverse exponential (green).

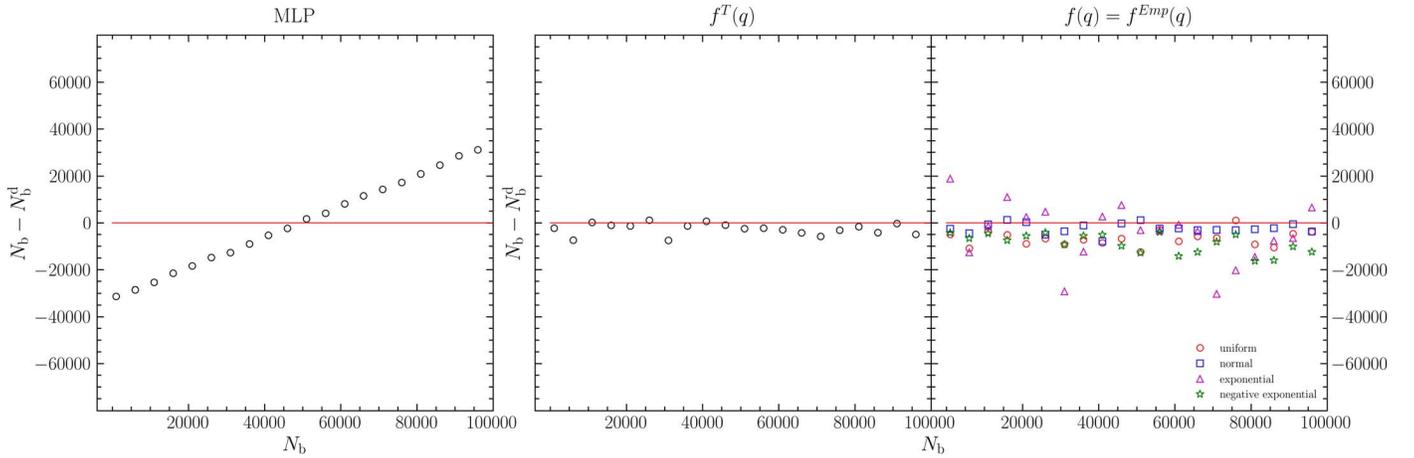

**Figure 16.** The relationship between the true number of binaries $N_b$ and the inferred number of binaries $N_b^d$ is examined in the observed sample without the NUV band. In the left column, the inferred number of binaries corresponds to the count obtained from our trained MLP model. The middle panel employs the exact same mass ratio distribution as the test sample to estimate the number of binaries, as described in Section 3.1.2. In the right column, in the absence of mass ratio distribution information, we assume a uniform mass ratio distribution to infer the number of binaries using the method outlined in Section 3.1.2. The symbols—open circle (red), open square (blue), open triangle (purple), and filled star (green)—represent uniform, normal, exponential, and negative exponential mass ratio distributions of the test samples, respectively.

Based on the number of binaries classified by the model and empirical mass ratio distribution, we have derived the binary number and binary fraction closer to the true values. The results show that the closer the given mass ratio distribution is to the true distribution, the more accurate the binary star number and fraction.

Once the CSST becomes operational, our method can be applied to identify potential candidates for double MS binaries and determine their binary fractions within the sample. We can then combine data from other photometric and spectroscopic surveys to accurately determine their mass, radius, age, and chemical composition, further understanding the relationship between the binary fraction and these physical parameters.

**Acknowledgments**

This research is funded by the National Natural Science Foundation of China (grants Nos. 12288102, 12125303, and 12090040/3), the National Key R&D Program of China (grant No. 2021YFA1600403), the Yunnan Fundamental Research Projects (grants No. 202201BC070003), the International Centre of Supernovae, Yunnan Key Laboratory (No. 202302AN360001), and the Yunnan Revitalization Talent Support Program–Science & Technology Champion Project (grant No. 202305AB350003). It is also supported by the China Manned Space Project under No. CMS-CSST-2021-A10.

**ORCID iDs**

Jia-jia Li ⓘ https://orcid.org/0009-0009-7824-5984
Jian-ping Xiong ⓘ https://orcid.org/0000-0003-4829-6245
Zhi-jia Tian ⓘ https://orcid.org/0000-0003-0220-7112
Chao Liu ⓘ https://orcid.org/0000-0002-1802-6917
Zhan-wen Han ⓘ https://orcid.org/0000-0001-9204-7778
Xue-fei Chen ⓘ https://orcid.org/0000-0001-5284-8001